\documentclass[twoside]{aa}

\usepackage{graphicx}
\usepackage{natbib}
\usepackage{latexsym}
\usepackage{amsmath}
\usepackage{amssymb}
\usepackage{amstext}
\usepackage{color}
\usepackage{psfrag}
\usepackage{threeparttable}
\usepackage{footnote}
\usepackage{float}
\makesavenoteenv{tabular}

\def\k{km s$^{-1}$}
\def\ks{km s$^{-1}$~}

\def\d{$^\circ$}
\def\m{$^\prime$}
\def\s{$^{\prime\prime}$}

\def\cm3{cm$^{-3}$}

\def\2{$^{12}$CO}
\def\3{$^{13}$CO}
\def\msol{M$_\odot$}
\def\lsol{L$_\odot$}

\begin{document}

\title{Outflowing activity in the UCHII region G045.47+0.05}
\author {M. E. Ortega \inst{1} \and S. Paron \inst{1,2} \and
  S. Cichowolski \inst{1} \and M. Rubio \inst{3} \and G. Dubner
  \inst{1} }

\institute{$^1$ Instituto de Astronom\'{\i}a y F\'{\i}sica del Espacio (IAFE),
             CC 67, Suc. 28, 1428 Buenos Aires, Argentina\\
            \email{mortega@iafe.uba.ar}\\
           $^2$ FADU - Universidad de Buenos Aires, Argentina\\
           $^3$ Departamento de Astronom\'ia, Universidad de Chile,
                Casilla 36-D, Santiago, Chile
}
\offprints{M. E. Ortega}

   \date{Received <date>; Accepted <date>}

\abstract {} {This work aims at investigating the molecular gas in the
  surroundings of the ultra-compact HII region G045.47+0.05 looking
  for evidence of molecular outflows.}{We carried out observations
  towards a region of 2\m$\times$2\m~centered at RA=19$^{\rm
    h}$14$^{\rm m}$25.6$^{\rm s}$, dec.=+11\d09\m27.6\s (J2000) using
  the Atacama Submillimeter Telescope Experiment (ASTE; Chile) in the
  $^{12}$CO J=3--2, $^{13}$CO J=3--2, HCO$^+$ J=4--3 and CS J=7--6
  lines with an angular resolution of 22\s. We complement these
  observations with public infrared data.} {We characterize the
    physical parameters of the molecular clump where G045.47+0.0 is
    embedded. The detection of the CS J=7--6 line emission in the
  region reveals that the ultra-compact HII region G045.47+0.0 has not
  completely disrupted the dense gas where it was born. The HCO$^+$
  abundance observed towards G045.47+0.0 suggests the presence of
  molecular outflow activity in the region. From the analysis of the
  $^{12}$CO J=3--2 transition we report the presence of bipolar
  molecular outflows with a total mass of about 300~\msol. We derive a
  dynamical time (flow's age) of about $10^5$ yr for the outflow
  gas, in agreement with the presence of an ultra-compact HII
  region. We identify the source 2MASS 19142564+1109283 as the massive
  protostar candidate to drive the molecular outflows. Based on
    the analysis of its spectral energy distribution we infer that it
    is an early B-type star of about 15~\msol. The results of this
  work support the scenario where the formation of massive stars, at
  least up to early B-type stars, is similar to that of low mass
  stars.}  {}

\titlerunning{Molecular outflows in G045.47+0.05}
\authorrunning{M. E. Ortega}


\maketitle

\section{Introduction}

The formation of high-mass stars remains one of the most significant
unsolved problems in astrophysics.  Despite the importance that
massive stars have in the structure and dynamic of the Galaxy, the
physical processes involved in their formation are less understood
than those of their low-mass counterpart.  Observationally, the main
problems arise from the fact that they are heavily obscured by dust,
are rare, and evolve very fast, making their detection very difficult.
The study of massive star formation also poses major theoretical
challenges because they begin burning their nuclear fuel and radiating
prodigious amounts of energy while still accreting. If the formation
of massive stars is similar to that of low mass stars (i.e. via
accretion from the surrounding envelope), a mass accretion rate
$\dot{M}$ of at least several orders of magnitude above the values
appropriate for the low mass star formation is required
\citep{tan02}.  At present, two theoretical scenarios are proposed
  to explain the formation of these stars: a monolithic collapse of
  turbulent gas on the scale of massive dense cores \citep{tan02},
  which is a scaled-up version of the low-mass star formation picture,
  and a competitive one where accretion occurs inside the
  gravitational potential of a cluster-forming massive dense core
  \citep{bon06}. This last model predicts that stars located near the
  center of the full gravitational potential accrete at much higher
  rates than do isolated stars.

The detection of molecular outflows associated with both, high and
low-mass young stellar objects, supports a scaled-up formation picture
\citep{beu02}.  In this context, the study of massive molecular
outflows and the associated driving source might contribute to discern
which is the scenario that prevails in a given star forming region.

It is well known that the generation of molecular outflows during the
formation of a high-mass star is a phenomenon that can take place even
when the UCHII region stage has been reached (\citealt{hunter1997};
\citealt{qin08}). In this work we report the study of the UCHII region
G045.47+0.05 (hereafter G45.47), through the analysis of its
associated molecular gas and searching for evidence of molecular
outflows and its associated driving source. G45.47 was first detected
by \citet{wood89} in radio continuum at 6~cm. This object is adjacent
to the extensively studied UCHII region G45.45+0.06 (hereafter
G45.45), which is part of a complex of five radio compact HII regions
(\citealt{mat77}; \citealt{giv05a}; \citealt{giv05b}). Such complex
and G45.47 are embedded in the molecular cloud GRSMC G045.49+00.04 (at
V$_{LSR} \sim$ 58~\k; \citealt{rat09}) and are located on the northern
border of the more extended HII region named G45L in
\citet{paron09}. Based on HI absorption profiles, \citet{kuc94}
derived a kinematics distance of 8.3~kpc for the UCHII region
G45.45. As G45.47 is part of the same complex, we adopt for this
object the same distance.

\citet{cas95} detected a class II CH$_3$OH maser emission at
  6.6~GHz at V$_{LSR}\sim56$~\k~ towards G45.47. Given that the
  6.6~GHz methanol maser is radiatively pumped by IR emission from the
  warm dust associated exclusively with massive young stellar object
  (MYSOs)\footnote{We define MYSOs to be young stellar objects (YSOs)
    that will eventually become main-sequence O or early B type stars
    (M$_*$ $\geq$ 8~\msol).} their detection is useful to study the
  kinematic of the gas and, in particular, to establish the systemic
  velocity \citep{cyga09}.

Based on high resolution molecular line observations towards G45.47,
\citet{wil96} identified five HCO$^+$ (1--0) clumps and suggested that
G45.47 is in the early stages of the formation of an OB cluster. From
an ammonia absorption study, \citet{hofner99} suggested the presence
of a remnant molecular core infalling onto the UCHII region. Later,
\citet{cyga08} identified an ``Extended Green Object'' (EGO) at the
position of G45.47, the EGO G45.47+0.05. Their identification as
  EGOs comes from the common coding of the 4.5~$\mu$m band as green in
  the three-color composite Infrared Array Camera (IRAC;
  \citealt{fazio04}) images from the {\it Spitzer} Telescope. Extended
  4.5~$\mu$m emission is thought to evidence the presence of shocked
  molecular gas in protostellar outflows. The association of EGOs with
  IRDCs and 6.7~GHz CH$_3$OH maser suggests that EGOs trace the
  formation of massive protostars. According to \citet{cyga08}, an
  EGOs is a MYSO with ongoing outflow activity and actively
  accreting.

In summary, G45.47 is a rich and complex region to study the formation
of a new generation of massive stars.  In this paper we present a new
study of the dense ambient medium where the UCHII region is evolving.
We investigate the molecular gas through several molecular lines
observed with the Atacama Submillimeter Telescope Experiment (ASTE;
Chile) and characterize the central source based on infrared public
data.

\section{Data}

\begin{figure*}
\centering
\includegraphics[width=17cm]{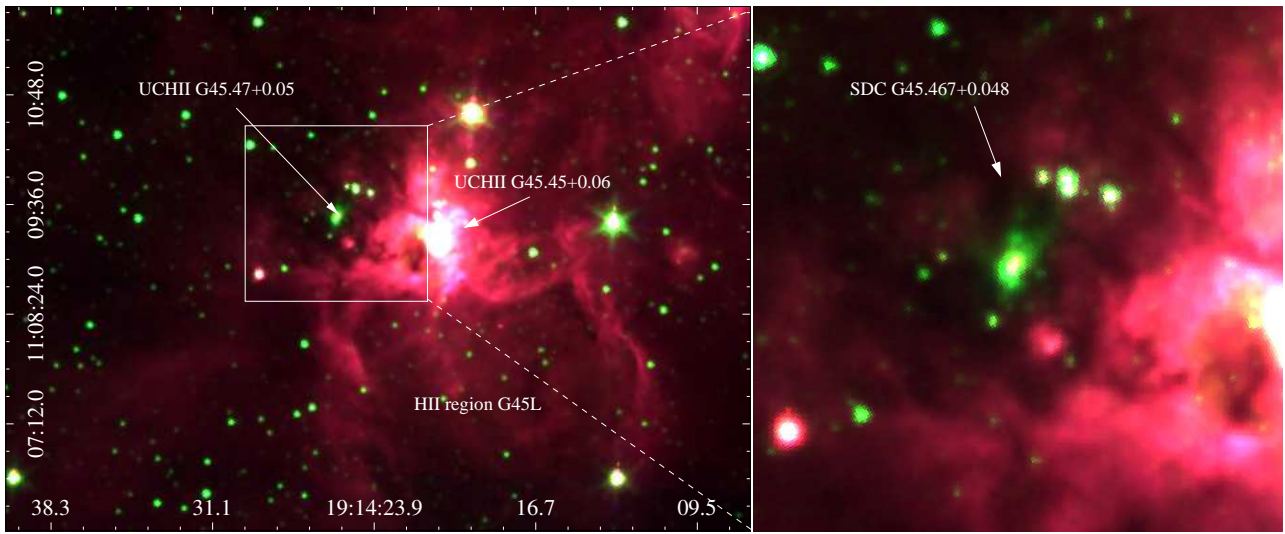}
\caption{{\it Spitzer}-IRAC three color images (3.6~$\mu$m = blue,
  4.5~$\mu$m = green and 8~$\mu$m = red). The white box in the left
  panel indicates the region mapped with ASTE. A close up view of this
  region is shown in the right panel. Red, green and blue scales go
  from 30 to 500, from 2 to 110, and from 1 to 80~MJy/sr,
  respectively.}
\label{intro}
\end{figure*}

The molecular line observations were carried out on June 12 and 13,
2011 with the 10~m Atacama Submillimeter Telescope Experiment
\citep[ASTE;][]{Ezawa04}. We used the CATS345~GHz band receiver, which
is a two-single band SIS receiver remotely tunable in the LO frequency
range of 324-372~GHz. We simultaneously observed $^{12}$CO J=3$-$2 at
345.796~GHz and HCO$^+$ J=4$-$3 at 356.734~GHz, mapping a region of
2\m $\times$2\m~ centered at RA = 19$^{\rm h}$14$^{\rm m}$25.6$^{\rm
  s}$, dec.=+11\d09\m27.6\s (J2000). We also observed $^{13}$CO
J=3$-$2 at 330.588~GHz and CS J=7$-$6 at 342.883~GHz towards the same
region. The mapping grid spacing was 20\s~in both cases and the
integration time was 20~sec ($^{12}$CO and HCO$^+$) and 40~sec
($^{13}$CO and CS) per pointing. All the observations were performed
in position switching mode. We verified that the off position (RA =
19$^{\rm h}$14$^{\rm m}$21.6$^{\rm s}$, dec.=+10\d59\m4\s, J2000) was
free of emission.  We used the XF digital spectrometer with a
bandwidth and spectral resolution set to 128~MHz and 125~kHz,
respectively. The velocity resolution was 0.11~\ks and the half-power
beamwidth (HPBW) was 22\s, for all observed molecular lines. The
system temperature varied from T$_{\rm sys}$ = 150 to 200~K. The main
beam efficiency was $\eta_{\rm mb} \sim$0.65. All the spectra were
Hanning smoothed to improve the signal-to-noise ratio. The baseline
fitting was carried out using second order polynomials for the
$^{12}$CO and $^{13}$CO transitions and fourth order polynomials for
the HCO$^+$ and CS transitions. The polynomia was the same for all
spectra of the map of a given transition.  The resulting rms noise of
the observations was about 0.2~K.  The data were reduced with
NEWSTAR\footnote{Reduction software based on AIPS developed at NRAO,
  extended to treat single dish data with a graphical user interface
  (GUI).}  and the spectra processed using the XSpec software
package \footnote{XSpec is a spectral line reduction package for
  astronomy which has been developed by Per Bergman at Onsala Space
  Observatory}.

The observations are complemented with near- and mid-IR data
extracted from public databases and catalogues, which are described in
the corresponding sections.

\section{Results and Discussion}

Figure \ref{intro}-(left) shows a composite {\it Spitzer}-IRAC
three-color image (3.6~$\mu$m = blue, 4.5~$\mu$m = green, and 8~$\mu$m
= red). The extended HII region G45L is centered at RA = 19$^{\rm
  h}$14$^{\rm m}$17$^{\rm s}$, dec.=+11\d07\m48\s (J2000) and is
delimited by two arclike structures observed at 8~$\mu$m.  The UCHII
region G45.47 (a.k.a. EGO G045.47+0.05) is the green source located at
RA=19$^{\rm h}$14$^{\rm m}$25.6$^{\rm s}$, dec.=+11\d 09\m 27.6\s
(J2000). The white box indicates the region mapped with ASTE. A zoom
up of this region (Fig. \ref{intro}-(right)) shows that the UCHII
region is embedded in the {\it Spitzer} dark cloud SDC G45.467+0.048
\citep{peretto09}. Infrared dark clouds (IRDCs) are dense molecular
clouds which appear as extinction features against the bright
mid-infrared Galactic background and have been suggested as the cold
precursors to high-mass stars \citep{rath06}.

\subsection{The molecular gas}
\label{molecular}

In this section, we present the molecular results starting the
  description with the transitions that trace the most inner part of
  the clump related to G45.47, and moving to those that map the
  external layers which may give information about the dynamic effects
  occurring in the gas.

\begin{figure*}
\centering
\includegraphics[width=14cm]{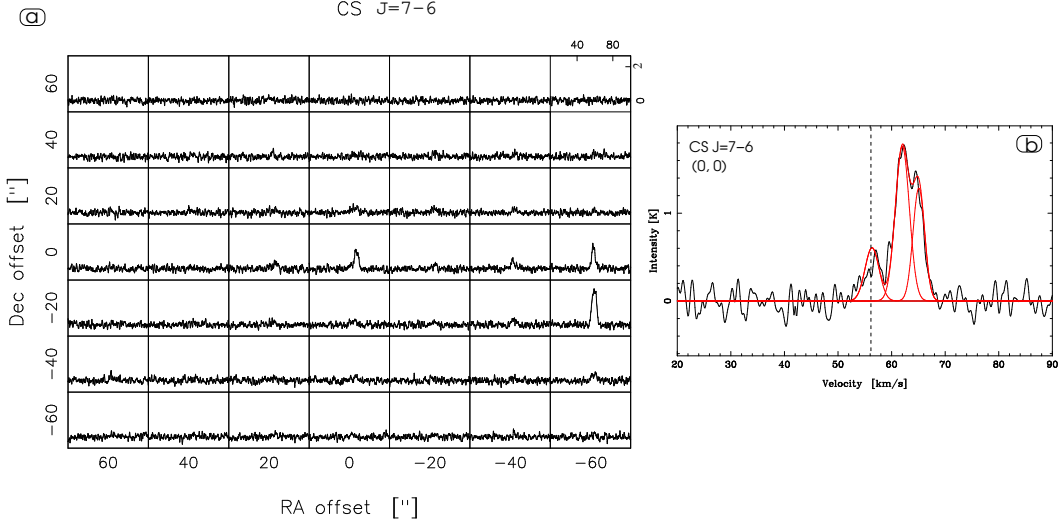}
\caption{a)CS J=7--6 spectra obtained towards the
  2\m~$\times$~2\m~region (white box in Fig. \ref{intro}-left) mapped
  with ASTE. b) Hanning smoothed profile of the CS J=7--6 line towards
  G45.47 at the position (0, 0).  The dashed line indicates the
  systemic velocity of the gas. The spectrum was deconstructed using
  three Gaussians, which are shown in red.}
\label{CSspectra}
\end{figure*}

Figure \ref{CSspectra}-a shows the CS J=7--6 spectra obtained towards
the 2\m~$\times$ 2\m~region (white box in Fig. \ref{intro}-left). The
mapped area includes G45.47 and part of the nearby UCHII region
G45.45, located at the (0, 0) and ($-$60, $-$20) offset,
respectively. As expected, the most prominent emission arises from
these two position, since the detection of the CS J=7--6 transition
reveals the presence of warm and dense gas. Figure \ref{CSspectra}-b
shows an enlargement of the spectrum observed at (0, 0). The emission
related to G45.47 shows a triple peak structure with velocity
components centered at about 56 (the systemic velocity of the gas),
62, and 65~\k (see Table \ref{gaussianfit}) with a pronounced dip at
about 59 and a less conspicuous one at $\sim $64~\k. The velocity
component centered at 56~\k is weak, with a peak about 3$\sigma$ of
the rms noise level. Since the probability of superposition along the
line of sight of more than one component in the CS J=7--6 line is very
low, we suggest that the dips reveal self-absorption effects in the
gas, consistent with an optically thick transition. Self-absorption
features demonstrate the existence of a density gradient in the clump
\citep{hira07} and is a common feature of optically thick lines in the
direction of embedded young stars. The depression discloses the
presence of relatively cold, foreground gas that absorbs photons from
warmer material behind it. The CS J=7--6 profile is asymmetric with
respect to the dip near v=59~\k as the redshifted component is
clearly brighter than the blueshifted one, suggesting that the
molecular gas is expanding. This is because in an expanding cloud a
line emission is composed by red and blueshifted photons, the
redshifted photons will encounter fewer absorbing material (which is
expanding outward) than would blueshifted photons and hence have
greater probabilities of escaping (e.g. \citealt{leu78};
\citealt{zhou92}; \citealt{leh97}).

\begin{figure*}
\centering \includegraphics[width=12cm]{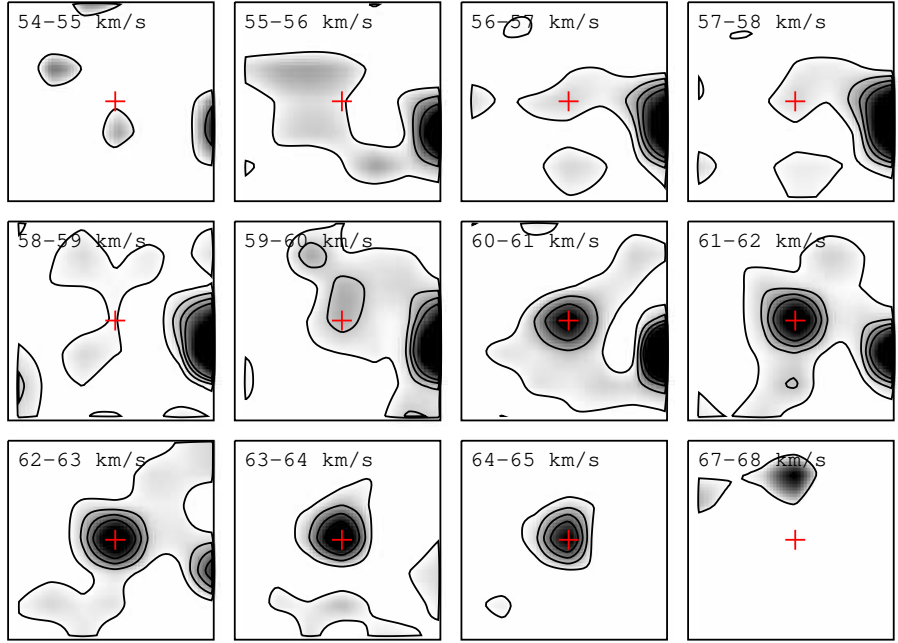}
\caption{Velocity channel maps of the CS J=7-6 emission averaged
    every 1~\k.  Greyscale goes from 0.15 (about 3$\sigma$ rms noise
    of an averaged channel map) up to 1.8 K. Contours levels are at
    0.15, 0.3, 0.45 and 0.6 K. The red cross represents the position
    of G45.47.}
\label{CS-maps}
\end{figure*}

Figure \ref{CS-maps} shows the velocity channel maps of the CS J=7--6
emission averaged every 1~\k. It can be noticed the presence of
molecular gas associated with G45.47 (red cross) in the velocity range
going from $\sim$ 55 to $\sim$65~\k. The other most conspicuous
molecular condensation, partially observed, corresponds to the gas
related to G45.45 and has a central velocity of $\sim$ 59~\k.

\begin{figure*}
\centering
\includegraphics[width=14cm]{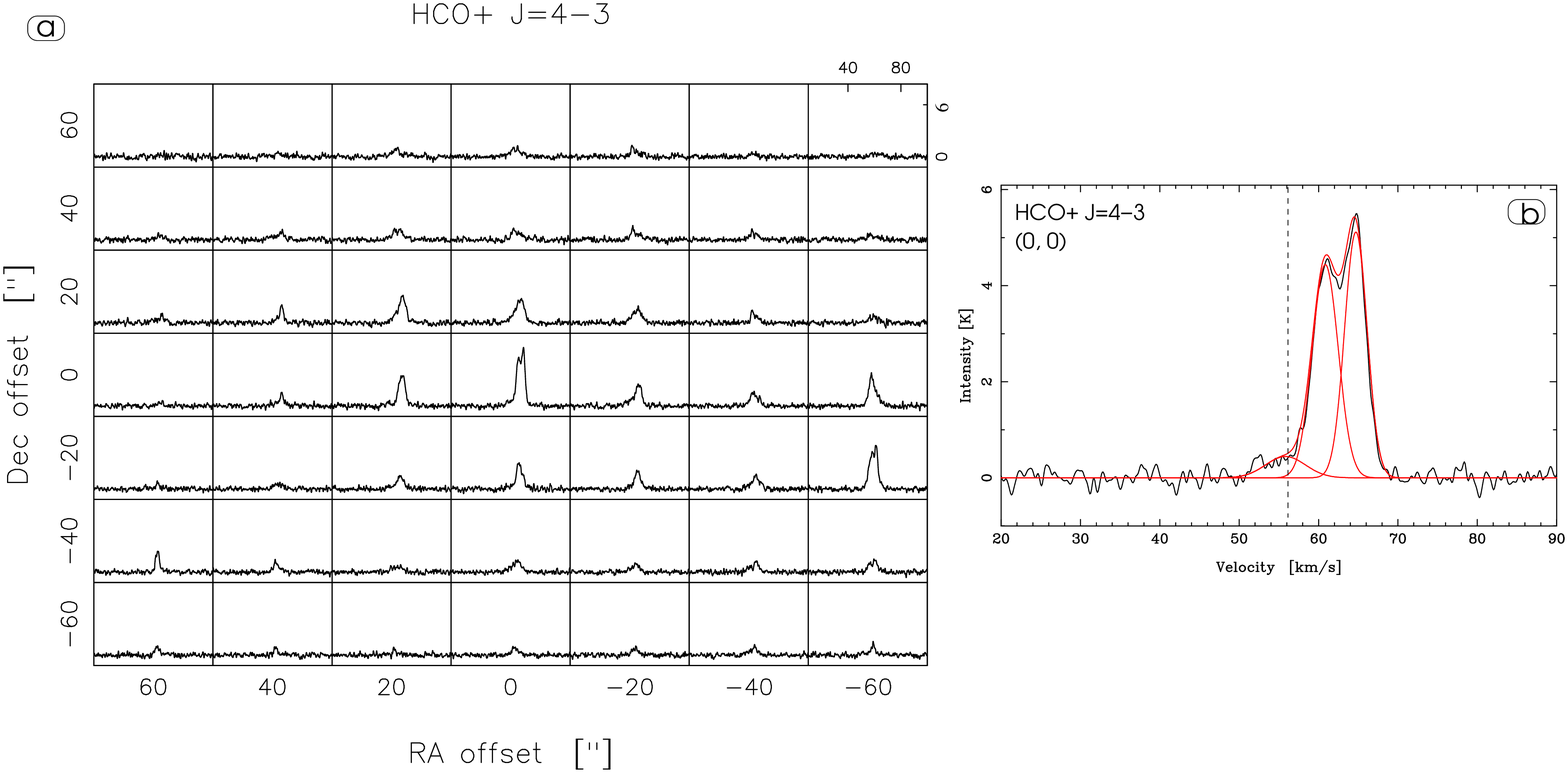}
\caption{a) HCO$^+$ J=4--3 spectra obtained towards the
  2\m~$\times$~2\m~region (white box in Fig. \ref{intro}-left) mapped
  with ASTE. b) Hanning smoothed profile of the HCO$^+$ J=4--3 line
  towards the position (0, 0) where it is G45.47. The dashed line
  indicates the systemic velocity of the gas. The spectrum was
  deconstructed using three Gaussians, which are shown in red.}
\label{HCO_spectra}
\end{figure*}

Figure \ref{HCO_spectra}-a shows the HCO$^+$ J=4--3 spectra obtained
towards the same region. As in the case of the CS J=7--6 spectrum it
can be seen that the HCO$^+$ emission towards (0, 0) can be
deconstructed in three velocity components, a very weak one centered
at about 56~\k~(intensity of about 2$\sigma$), plus two brighter ones
at 61 and 65~\k, with only one dip centered at $\sim$ 63~\k.

From Fig. \ref{HCO_spectra}-a it can be noticed that the HCO$^+$
spectrum on position ($-$60, $-$20), that is at the position of the
UCHII region G45.45, has also a dip at about 60~\k~similarly to what
it is observed in the CS J=7--6 line towards the same position.

The HCO$^+$ spectra also show evidence of emission near position (60,
$-$40). This molecular feature is in positional coincidence with an
infrared (IR) source located at RA=19$^{\rm h}$14$^{\rm m}$27.7$^{\rm
  s}$, dec.=+11\d08\m33\s (J2000) (see Fig \ref{intro}). As this
HCO$^+$ spectrum has the same kinematic structure than the spectrum
towards ($-$60, $-$20) including the presence of a dip at the same
velocity of about 60~\k, we can conclude that this IR source is
embedded in a molecular filament that must be connected with G45.45.

\begin{figure*}
\centering
\includegraphics[width=12cm]{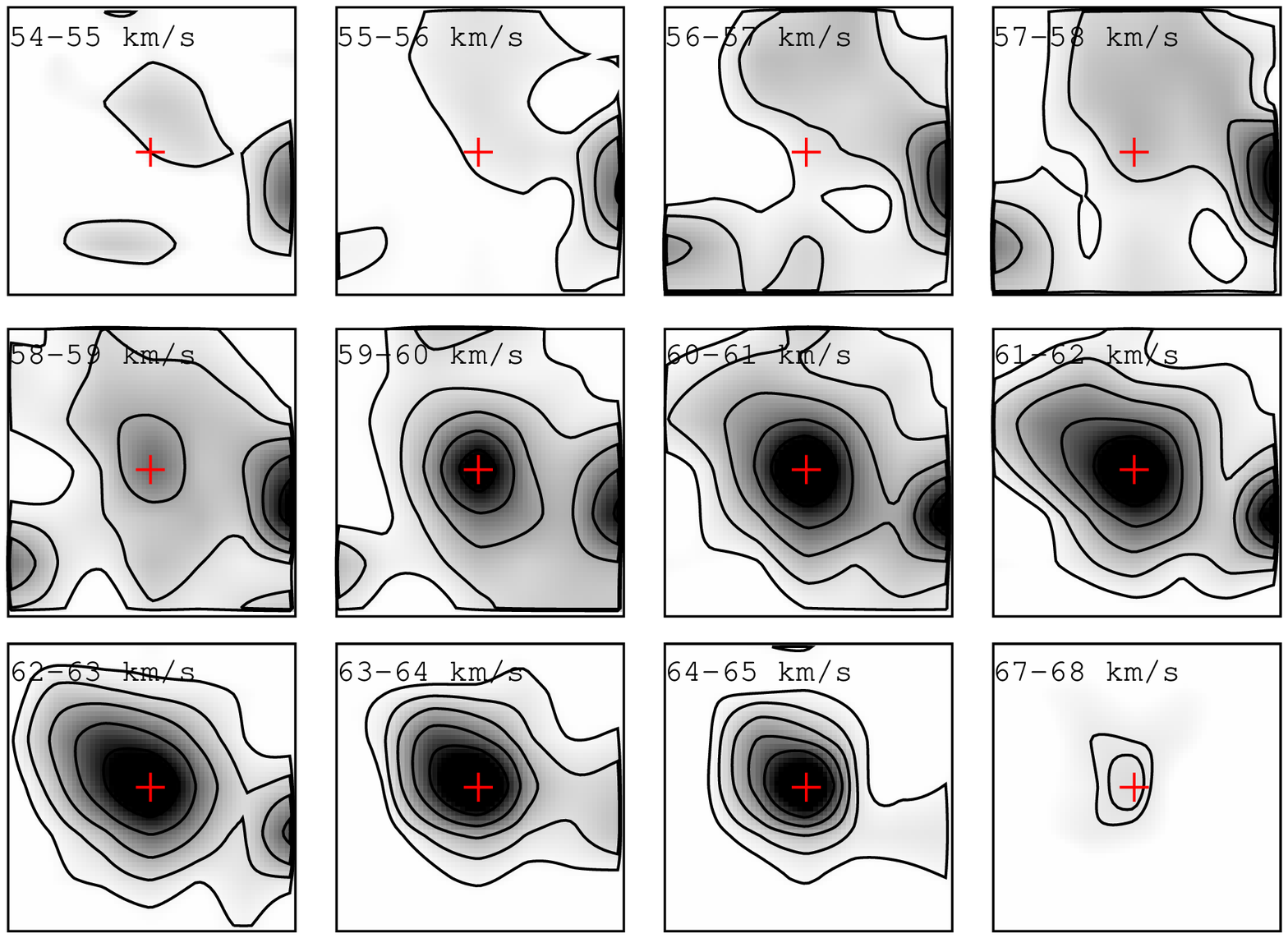}
\caption{Velocity channel maps of the HCO$^+$ J=4--3 emission averaged
  every 1~\k. Greyscale goes from 0.15 (about 3$\sigma$ rms noise of
  an averaged channel map) to 2.5~K. Contours levels are at 0.15, 0.5,
  0.8, 1.3, and 1.9~K. The red cross represents the position of
  G45.47.}
\label{HCO-maps}
\end{figure*}

Figure \ref{HCO-maps} shows the velocity channel maps of the HCO$^+$
J=4--3 emission averaged every 1~\k. The HCO$^+$ J=4--3 emission
related to G45.47 is visible between $\sim$ 55 and $\sim$ 68~\k, while
the emission associated with G45.45 goes from $\sim$ 54 to $\sim$
63~\k, respectively. Between $\sim$ 56 and $\sim$ 60~\k part of the
molecular condensation related to the IR source mentioned above can be
appreciated.

\begin{figure*}
\centering \includegraphics[width=15cm]{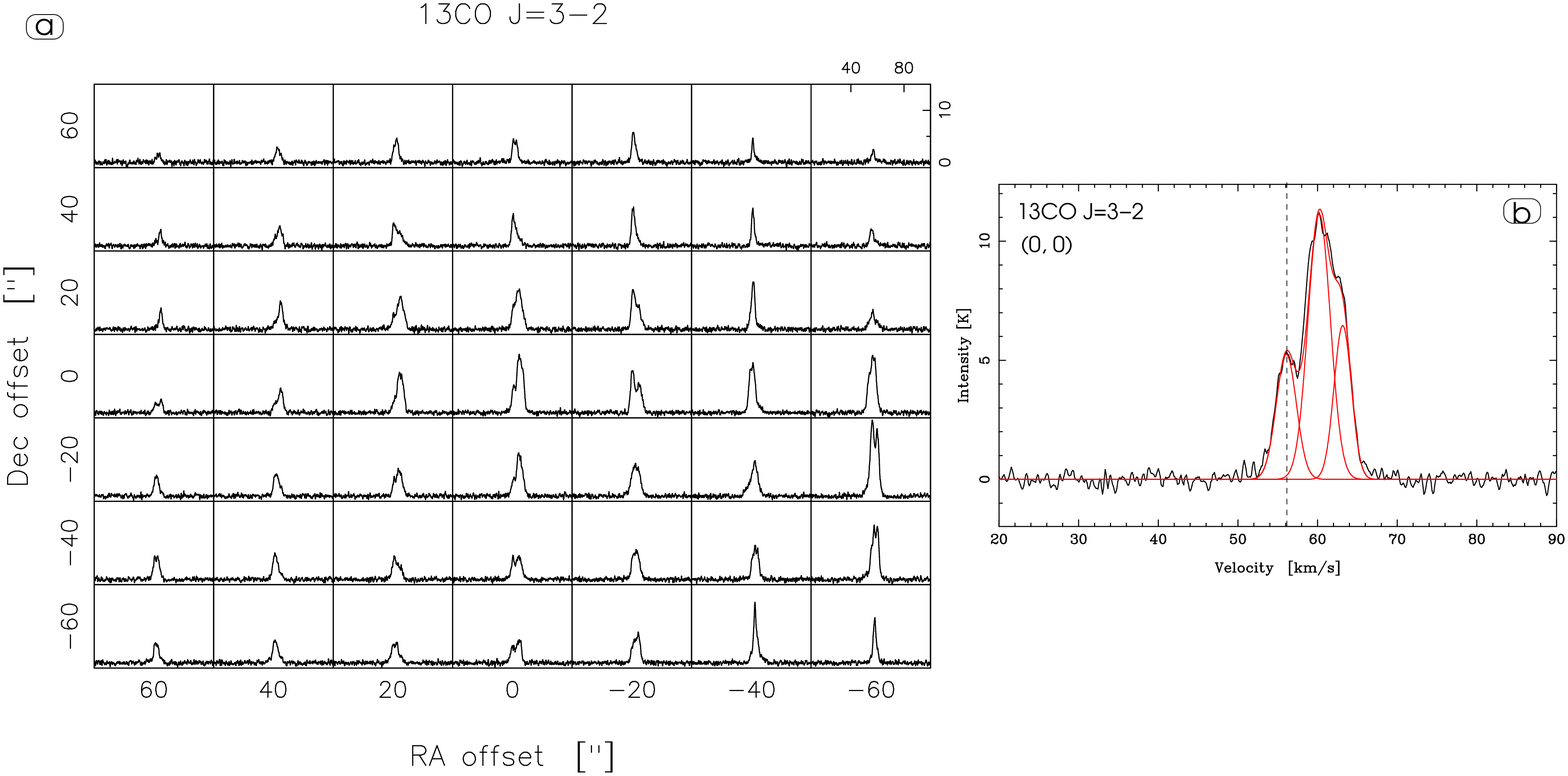}
\caption{$^{13}$CO J=3--2 spectra obtained towards the
  2\m~$\times$~2\m~region (white box in Fig. \ref{intro}-left) mapped
  with ASTE. b) Hanning smoothed profile of the $^{13}$CO J=3--2
    line towards G45.47.  The dashed line indicates the systemic
  velocity of the gas. The spectrum was deconstructed using three
  Gaussians, which are shown in red.}
\label{13CO_spectra}
\end{figure*}

\begin{figure*}
\centering \includegraphics[width=12cm]{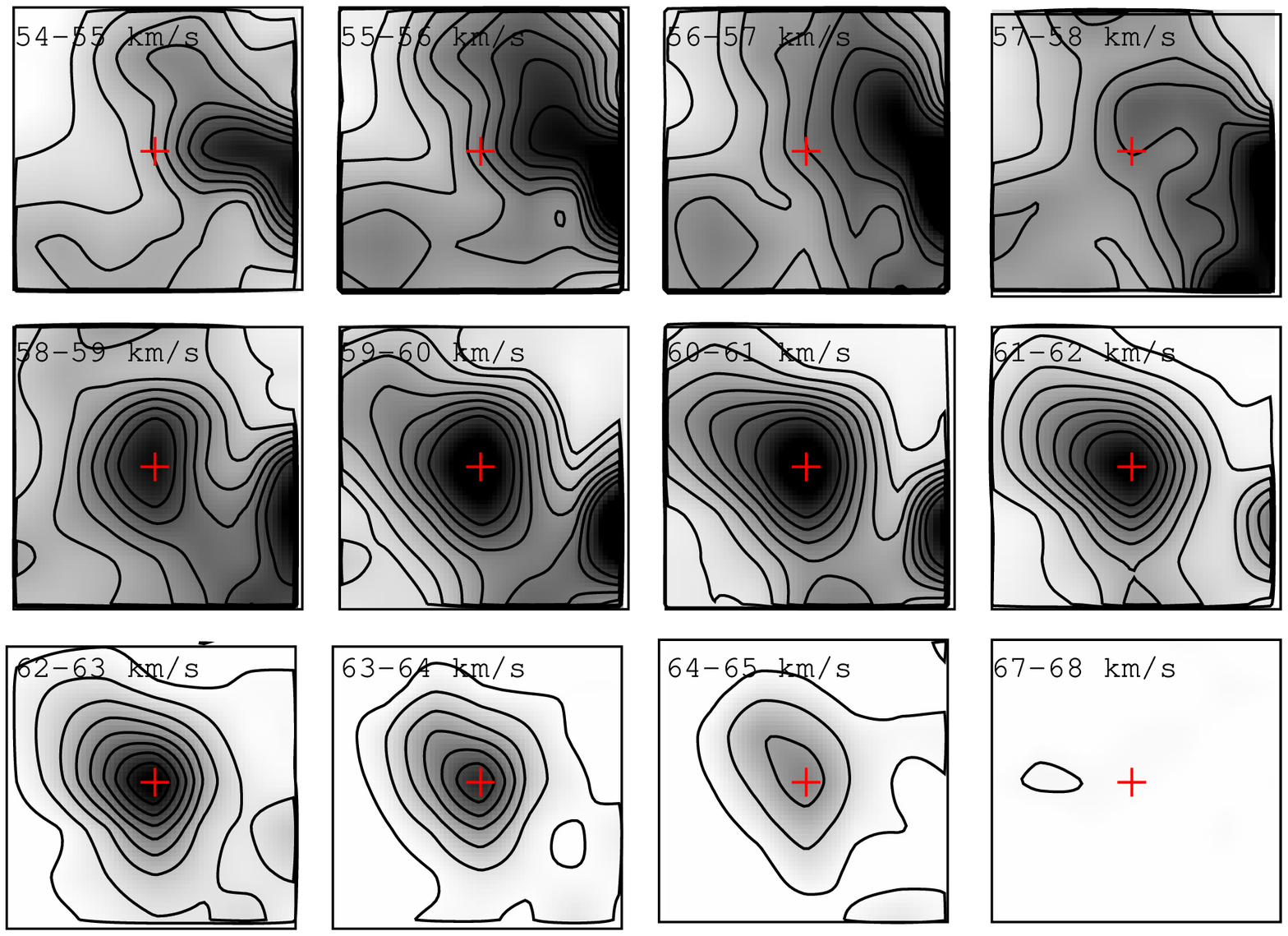}
\caption{Velocity channel maps of the $^{13}$CO J=3--2 emission
  averaged every 1~\k.  Greyscale goes from 0.15 up to 9~K. Contours
  levels are at 0.6, 2, 4, 5, 6, 7, 8, and 9~K. The red cross
  represents the position of G45.47.}
\label{13CO-maps}
\end{figure*}

Figure \ref{13CO_spectra} shows the $^{13}$CO J=3--2 spectra towards
the same region.  The $^{13}$CO emission towards (0, 0) can be
decomposed in three velocity components centered at about 56 (the
systemic velocity of the gas), 60, and 63~\k, while only one dip
centered at $\sim$ 58~\k can be appreciated. The depression at
58~\k~can be observed in almost all the $^{13}$CO J=3--2 spectra in
the region.  {\bf Figure \ref{13CO-maps} shows the velocity channel
  maps of the $^{13}$CO J=3--2 emission averaged every 1~\k.}

\begin{figure*}
\centering \includegraphics[width=14cm]{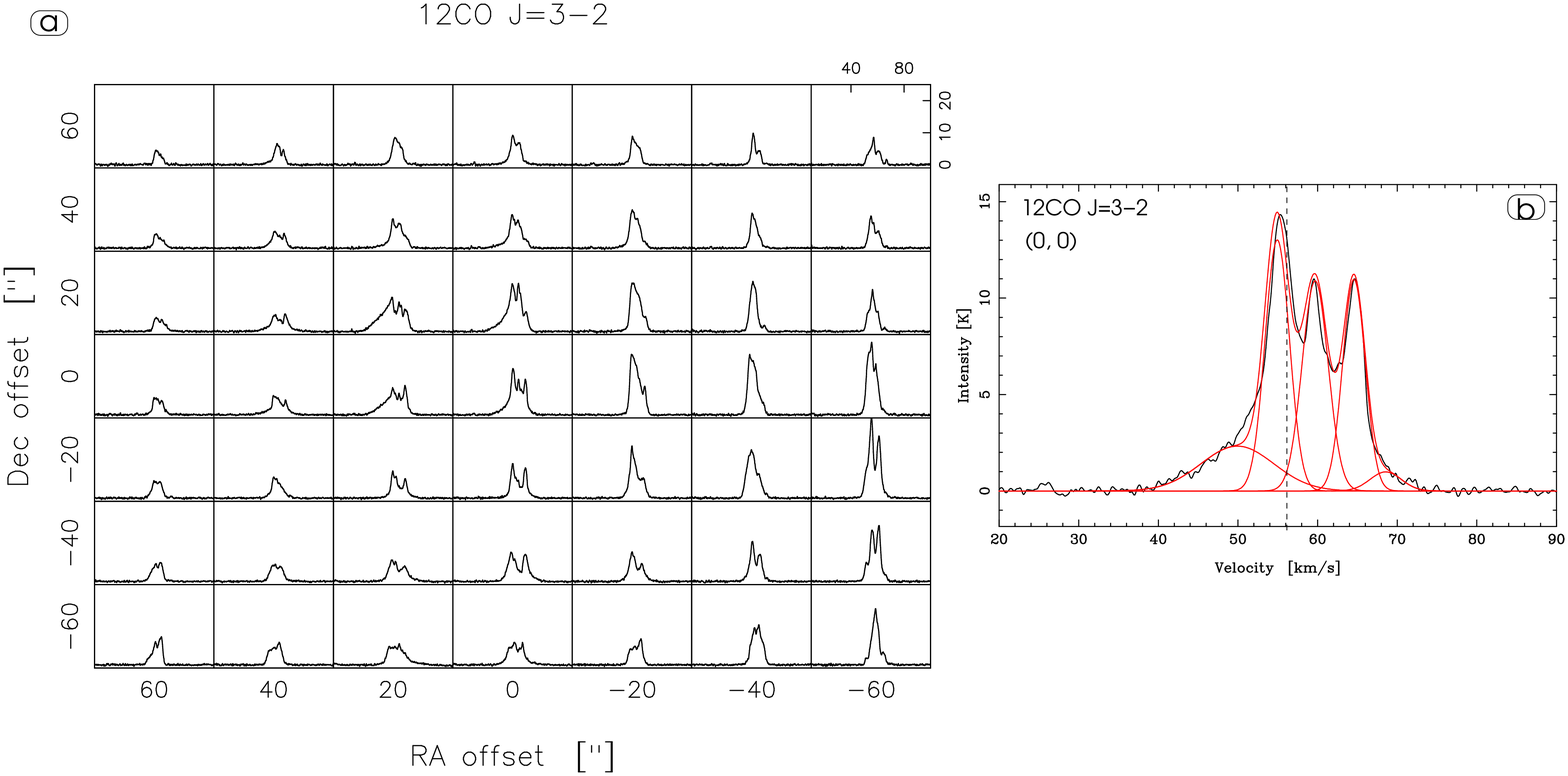}
\caption{a) $^{12}$CO J=3--2 spectra obtained towards the
  2\m~$\times$~2\m~region (white box in Fig. \ref{intro}-left) mapped
  with ASTE. b) Hanning smoothed profile of the $^{12}$CO J=3--2 line
  towards G45.47. The dashed line indicates the systemic velocity of
  the gas. The spectrum was deconstructed using five Gaussians, which
  are shown in red.}
\label{12CO_spectra}
\end{figure*}

The $^{12}$CO J=3--2 spectra (Fig. \ref{12CO_spectra}) exhibit a more
complicated behaviour. The profile towards the position (0, 0) has a
triple peak structure with components centered at about 55, 60, and
64~\k and two dips at about 58 and 62~\k. This spectrum is broadened,
suggesting the presence of outflowing activity in the region with the
blue wing centered near the position (20, 20) and the red wing around
the (0, $-$40) offset.

\begin{figure*}
\centering
\includegraphics[width=12cm]{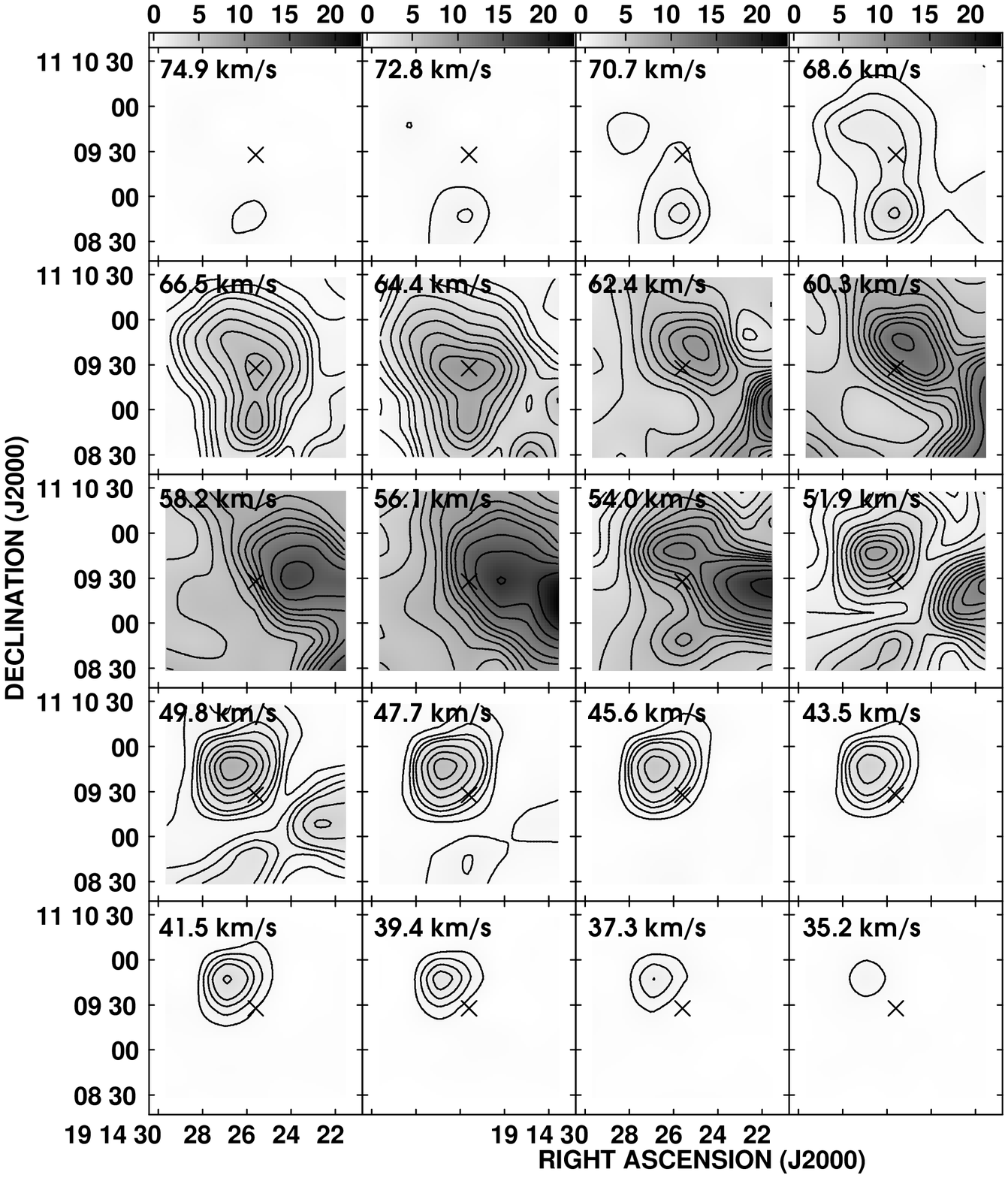}
\caption{Velocity channel maps of the $^{12}$CO J=3--2 emission
    averaged every 2.1~\k. The given velocities correspond to the
    higher velocity of each interval. Greyscale goes up to
    23~K. Contours are above the 5$\sigma$ of the rms noise level. The
    cross indicates the position of G45.47.}
\label{12CO_maps}
\end{figure*}

In Figure \ref{12CO_maps} we show the velocity channel maps of the
$^{12}$CO J=3--2 emission averaged every 2.1~\k. Among all the
observed molecular condensations, we draw the attention onto the
clumps related to G45.47 and G45.45. The molecular clump related to
G45.47 is observed between $\sim$ 62 and 66~\k and is seen slightly
shifted to the northwest between $\sim$ 58 and 62~\k. The clump
associated with G45.45 (partially observed) is centered at RA=19$^{\rm
  h}$14$^{\rm m}$22$^{\rm s}$, dec.=+11\d09\m20\s (J2000) in the
velocity interval going from 47 to 63~\k. Although both molecular
condensations have associated different velocity ranges, they are
connected through the extended emission shown in the velocity interval
going from 50 to 65~\k. The molecular gas related to the spectral
wings appears as two conspicuous molecular features centered at
RA=19$^{\rm h}$14$^{\rm m}$27$^{\rm s}$, dec.=+11\d09\m45\s (J2000)
between 35 and 53~\k~ and at RA=19$^{\rm h}$14$^{\rm m}$26$^{\rm s}$,
dec.=+11\d08\m45\s (J2000) between 65 and 76~\k.  These features will
be further discussed in Section \ref{outflows}.

The four transitions have, within errors, the same main velocity
components at about 55, 60, and 64~\k~towards the position (0, 0). By
the other hand, the velocity components observed at about 51 and
68~\k~in the $^{12}$CO J=3--2 spectrum are not detected in the other
three lines. Table \ref{gaussianfit} lists the emission peaks
parameters derived from a Gaussian fitting for the four molecular
transitions on the position (0, 0).  T$_{mb}$ represents the peak
brightness temperature and V$_{LSR}$ the central velocity referred to
the Local Standard of Rest. Errors are formal 1$\sigma$ value for the
model of Gaussian line shape. All the spectra towards the (0,0)
position have the same self-absorption dip at about 58-59~\k, which
correspond to the central velocity of the molecular cloud GRSMC
045.49+00.04 where G45.47 is embedded. Even more, this spectral
feature is observed in all transitions towards the region G45.45 and
the IR source located at RA=19$^{\rm h}$14$^{\rm m}$27.7$^{\rm s}$,
dec.=+11\d08\m33\s (J2000) (see Fig \ref{intro}).

\begin{figure*}
\centering \includegraphics[width=13cm]{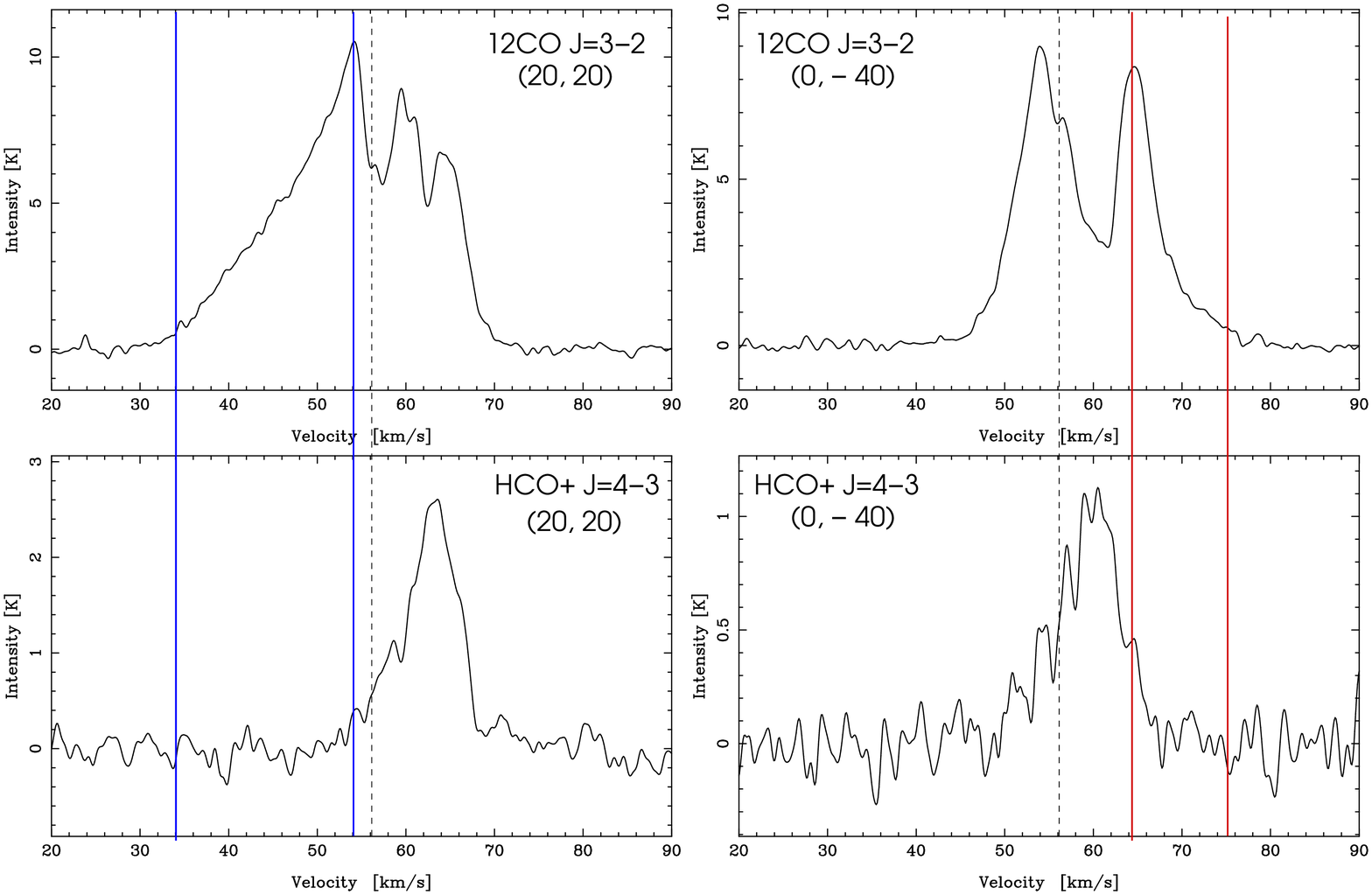}
\caption{Comparison between the $^{12}$CO J=3--2 and the HCO$^+$
  J=4--3 spectra on the positions (20, 20) and (0, -40). The dashed
  line indicates the systemic velocity. The velocity intervals
  corresponding to the blue and red wings are shown. The four spectra
  were Hanning smoothed.}
\label{spectralwings}
\end{figure*}

\begin{table} 
\caption{Emission peaks parameters derived from a Gaussian fitting for the four molecular transitions spectra on
  the position (0, 0).} \centering
\begin{tabular}{cccc}
\hline 
Transition & V$_{LSR}$ [\k] &  T$_{mb}$ [K] & $\Delta$v [\k]\\
\hline 
CS J=7--6 & 56.3$\pm$0.9 & 0.6$\pm$0.2 & 3.1$\pm$0.6\\
          & 62.0$\pm$0.8 & 1.8$\pm$0.1 & 3.1$\pm$0.7\\
          & 64.8$\pm$1.1 & 1.3$\pm$0.1 & 2.5$\pm$0.9\\
\hline
HCO$^+$ J=4--3  & 55.8$\pm$1.2 & 0.4$\pm$0.2 & 5.9$\pm$0.5\\
                & 60.8$\pm$1.4 & 4.4$\pm$0.8 & 3.8$\pm$0.6\\
                & 64.7$\pm$1.7 & 5.1$\pm$0.6 & 3.3$\pm$0.7\\
\hline
$^{13}$CO J=3--2 & 55.9$\pm$0.6 & 5.2$\pm$0.4 & 2.9$\pm$0.4\\
                & 60.5$\pm$0.5 & 11.1$\pm$0.8 & 3.3$\pm$0.6\\
                & 63.3$\pm$0.4 & 6.4$\pm$0.7 & 2.6$\pm$0.7\\
\hline
$^{12}$CO J=3--2 & 50.7$\pm$0.3 & 3.0$\pm$0.2 & 9.0$\pm$0.3\\  
                & 55.2$\pm$0.1 & 13.1$\pm$1.1 & 3.3$\pm$0.4\\
                & 59.8$\pm$0.2 & 10.6$\pm$0.8 & 3.6$\pm$0.6\\
                & 64.4$\pm$0.6 & 10.6$\pm$0.7 & 3.3$\pm$0.5\\
                & 68.1$\pm$0.4 & 1.3$\pm$0.2 & 4.4$\pm$0.4\\
\hline

\label{gaussianfit}
\end{tabular}
\end{table}

\subsection{Column density and mass estimates of the molecular clump associated with G45.47}
\label{clump}

We estimate the $^{13}$CO J=3--2 opacity, $\tau_{13}$, based on the
following equation:

\begin{equation}
\small 
\tau_{13}=-ln\left(1-\frac{T_{peak}(^{13}{\rm CO})}{T_{peak}(^{12}{\rm CO})}\right)
\end{equation}

\noindent where we consider T$_{peak}$ from the position (0, 0). We
obtain, $\tau_{13} \sim$ 1.9 which reveals that the $^{13}$CO J=3--2
emission is optically thick towards G45.47, in agreement with the
observed profile towards (0,0) offset (see
Fig. \ref{13CO_spectra}). 

The excitation temperature, $T_{ex}$, of the $^{13}$CO J=3--2 line was
estimated from:

\begin{equation}
\small 
^{13}T_{peak}=\frac{h\nu}{k}
\left(\frac{1}{e^{h\nu/k T_{ex}}-1}-\frac{1}{e^{h\nu/k T_{BG}}-1}\right)
\times (1-e^{-\tau_{13}})
\end{equation}

\noindent where for this line $h\nu/k=15.87$. Assuming $T_{BG}$ =
2.7~K, and considering the peak brightness temperature for the
$^{13}$CO J=3--2 at (0, 0) offset, $^{13}T_{peak}$ = 11.15~K, we
derive a $T_{ex} \sim$ 20~K for the $^{13}$CO J=3--2 line.

Finally, using the
RADEX \footnote{http://www.sron.rug.nl/$\sim$vdtak/radex/radex.php}
code \citep{tak07} we derive the $^{13}$CO J=3--2 column density and
the H$_2$ volume density. The RADEX model uses the mean escape
probability approximation for the radiative transfer equation.

Adopting $T_{ex} \sim$ 20~K, $\tau_{13} \sim$ 1.9, and $^{13}T_{peak}$
= 11.15~K, we obtain a $^{13}$CO column density N($^{13}$CO) $\sim 2.8
\times 10^{17} {\rm cm}^{-2}$ and n(H$_2$) $\sim 10^5 {\rm cm}^{-3}$.
Considering an abundance ratio of [H$_2$]/[$^{13}$CO] = 77$\times
10^4$ \citep{wil94} we estimate the H$_2$ column density, N(H$_2) \sim
2.1 \times 10^{23}{\rm cm}^{-2}$. Finally, using the relation $M=\mu
m_Hd^2\Omega {\rm N(H_2)}$, where $\mu$ is the mean molecular weight
per H$_2$ molecule ($\mu \sim 2.72$), $m_H$ the hydrogen atomic mass,
$d$ the distance, and $\Omega$ the solid angle subtended by the
structure, we calculate the total mass of the clump in $M \sim 10^4$
\msol.
 
As an independent estimate, we derive the beam-averaged gas column
density, the mass, and the volume density of the clump based on the
dust continuum emission. In particular, we use the integrated flux of
the continuum emission at 1.1~mm as obtained from The Bolocam Galactic
Plane Survey II Catalog (BGPS II; \citealt{ros10}). The 1.1~mm
continuum emission is mostly originated in optically thin dust
\citep{hild83}. Following \citet{be02} and \citet{hild83} we calculate
the mass and the gas column density of the clump using:

\begin{eqnarray}
M_{gas}=\frac{1.3 \times
  10^{-3}}{J_{\nu}(T_{dust})}\frac{a}{0.1 \mu{\rm m}}\frac{\rho}{3
  {\rm g cm}^{-3}}\frac{R}{100}\frac{F_{\nu}}{{\rm
    Jy}}\nonumber\\\left(\frac{d}{{\rm kpc}}\right)^2 \left(\frac{\nu}{2.4 {\rm
    THz}}\right)^{-3-\beta} [{\rm M}_{\odot}]
\end{eqnarray}

\noindent and 

\begin{eqnarray}
N_{gas}=\frac{7.8 \times 10^{10}}{J_{\nu}(T_{dust})\Omega_b}\frac{a}{0.1
  \mu{\rm m}}\frac{\rho}{3 {\rm g
    cm}^{-3}}\frac{R}{100}\frac{F_{\nu}}{{\rm
    Jy}}\nonumber\\ \left(\frac{\nu}{2.4 {\rm THz}}\right)^{-3-\beta}
[{\rm cm}^{-2}]
\end{eqnarray}

\noindent where $J_{\nu}(T_{dust}$) = [exp($h\nu/kT_{dust})-1]^{-1}$
and $\Omega_b, a, \rho, R,$ and $\beta$ are the beam solid angle,
grain size, grain mass density, gas-to-dust ratio, and grain
emissivity index for which we used the values of (33\s)$^2$ in
radians, 0.1~$\mu$m, 3~g cm$^{-3}$, 100, and 2, respectively
(\citealt{hun97}, \citealt{hun00}, and \citealt{mol00}). Based on the
work of \citet{sri02} who derived dust temperatures ranging between 30
and 60~K for a sample of several massive star forming regions, we
adopt $T_{dust}$ = 45~K. For a distance $d$ = 8.3~kpc and an
integrated flux intensity $F_{\nu}$ = 5.18~Jy at 1~mm \citep{ros10} we
obtain $N_{gas} \sim 4 \times 10^{23}$~cm$^{-2}$, $M_{gas} \sim
8520$~\msol, and a volume density, $n$(H$_2) \sim 1.4 \times
10^5$~cm$^{-3}$.  These values are in good agreement with those
derived from the $^{13}$CO J=3--2 line using RADEX.

\subsection{Molecular outflows associated with G45.47}
\label{outflows}

As discussed in Section \ref{molecular}, the presence of spectral
wings in the $^{12}$CO J=3--2 spectrum obtained towards G45.47 is a
strong indicator that molecular outflow activity is taking place in
the region. To characterize the associated molecular outflows it is
first necessary to separate the outflowing gas from the molecular
material of the clump, identifying the velocity ranges related to each
structure.

We consider two independent methods to determine these velocity
ranges. The first method consists in the identification of the blue
and red spectral wings based on a comparison between the $^{12}$CO
J=3--2 and HCO$^+$ J=4--3 spectra. Figure \ref{spectralwings} shows
the comparison between both spectra on positions (20, 20) and (0,
$-$40) where the blue and red wings are largest.  Considering emission
up to about 2~$\sigma$ of rms noise level, it can be noticed a blue
and a red wing in the $^{12}$CO spectra between $\sim$ 34 and $\sim$
54 and between $\sim$ 64 and $\sim$ 75~\k, respectively.

\begin{figure}[h]
\centering
\includegraphics[width=8cm]{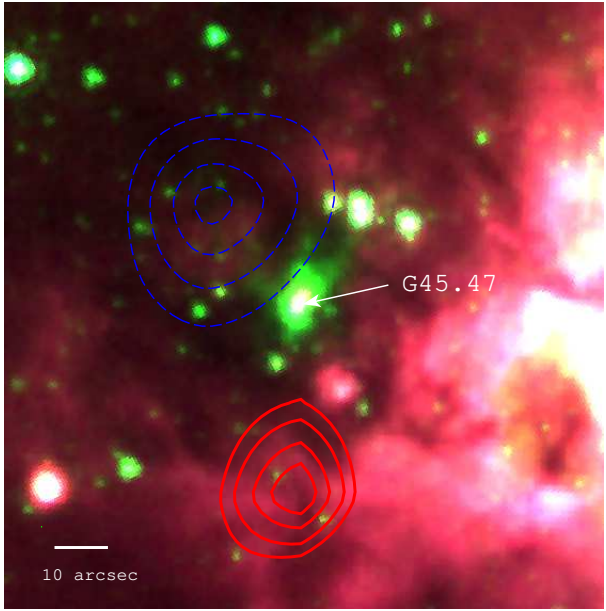}
\caption{{\it Spitzer}-IRAC three color image (3.6~$\mu$m = blue,
  4.5~$\mu$m = green and 8~$\mu$m = red) of G45.47. The blue and red
  contours represent the $^{12}$CO J=3--2 emission integrated from 34
  to 54~\k~(blue lobe), and from 64 to 75~\k~(red lobe),
  respectively. The blue contours are at 400, 500, 600, and 670 K~\k~
  and the red ones are at 140, 160, 180, and 200 K~\k.}
\label{blue-red-wings}
\end{figure}

The second method to identify the molecular outflows is based on the
inspection of the $^{12}$CO J=3--2 data cube, channel by channel,
trying to spatially separate both outflows's lobes. In Section
\ref{molecular} we mentioned two molecular features in the $^{12}$CO
J=3--2 emission centered at RA=19$^{\rm h}$14$^{\rm m}$27$^{\rm s}$,
dec.=+11\d09\m45\s (J2000) between $\sim$ 34 and $\sim$ 54~\k~and at
RA=19$^{\rm h}$14$^{\rm m}$26$^{\rm s}$, dec.=+11\d08\m45\s (J2000)
between $\sim$ 64 and $\sim$ 75~\k~which we identify as the outflow's
lobes related to G45.47 (see Fig. \ref{12CO_maps}). These molecular
features can be identified as the spectral wings detected in the
$^{12}$CO J=3--2 spectra through the first method.  Figure
\ref{blue-red-wings} shows a {\it Spitzer}-IRAC three color image
(3.6~$\mu$m = blue, 4.5~$\mu$m = green and 8~$\mu$m = red) of
G45.47. The blue and red contours represent the $^{12}$CO J=3--2
emission integrated from $\sim$ 34 to 54~\k~(blue lobe) and from
$\sim$ 64 to 75~\k~(red lobe), respectively.

Following \citet{sco86} we estimate the $^{12}$CO optical depth,
$\tau_{12}$, of the gas in the molecular outflows using: 

\begin{equation}
\small
\frac{^{12}{\rm T}_{mb}}{^{13}{\rm T}_{mb}}=\frac{1-exp(-\tau_{12})}{1-exp(-\tau_{12}/X)}
\end{equation}

\noindent where $X$=[$^{12}$CO]/[$^{13}$CO] is the isotope abundance
ratio. Using the relation $X=6.21 \times D_{GC} + 18.71$
\citep{mil05}, where D$_{GC}$ = 6.1~kpc is the distance between G45.47
and the Galactic Center, we obtain an isotope abundance ratio $X$= 57
for this region. We adopt a constant $X$ throughout the outflows
\citep{cab90}. Since we could not identify any blue wing in the
$^{13}$CO spectrum on position (20, 20), the
$^{12}$T$_{mb}/^{13}$T$_{mb}$ ratio for both wings was estimated
considering only the red one. We obtain a
$^{12}$T$_{mb}/^{13}$T$_{mb}$ ratio of about 4.2 and derive a
$^{12}$CO optical depth of $\tau_{12} \sim$ 15, which will be adopted
for both spectral wings in further calculations.

We can now estimate the $^{12}$CO column density of both outflow lobes
from (see e.g. \citealt{buc10}):

\begin{equation} 
\small 
{\rm N}(^{12}{\rm CO})=7.96\times10^{13}
e^{\frac{16.6}{T_{ex}}}\frac{T_{ex}+0.92}{1-exp(\frac{-16.6}{T_{ex}})}
\int \tau_{12} {\rm dv} 
\end{equation}

\noindent Taking into account that the $^{12}$CO J=3--2 is an
optically thick transition ($\tau \geq 1$), the integral can be
approximated by:

\begin{equation}
\small \int \tau_{12} {\rm dv} =
\frac{1}{J(T_{ex})-J(T_{BG})}\frac{\tau_{12}}{1-e^{-\tau_{12}}}\int
     {\rm ^{12}T}_{mb}~{\rm dv}
\label{thick}
\end{equation}

\noindent By integrating the $^{12}$CO emission from 34 to 54~\k~and
from 64 to 75~\k~we obtain N(CO)$_{blue} \sim 5.2 \times 10^{17} {\rm
  cm}^{-2}$ and N(CO)$_{red} \sim 2 \times 10^{17} {\rm cm}^{-2}$,
respectively. The mass of each wing can be derived from the relation
$M=\mu m_Hd^2\Omega X({\rm CO})^{-1}{\rm N(CO)}$, where $X$(CO) is the
$^{12}$CO relative abundance to H$_2$ ($X$(CO)$\sim 7.4 \times
10^{-5}$). We obtain M$_{blue} \sim 300$ \msol~and M$_{red} \sim 120$
\msol, yielding a total outflow mass of about 420~\msol. In this way,
the estimated total outflows mass represents about 4\% of the clump
mass, which is in agreement with the results of \citet{beutPhD02} who
established that approximately 4\% of the clump gas is entrained in
the molecular outflows.

The momentum and the kinetic energy of the wings can be derived from
$P = M V_c$ and $E_k = 0.5 M V_c^2$, where $V_c$ is a characteristic
velocity estimated as the difference between the maximum velocity of
detectable $^{12}$CO emission in the wing and the systemic velocity of
the gas ($\sim$ 56~\k). Taking into account a $V_c^{blue} \sim$ 22~\k~
and a $V_c^{red} \sim$ 19~\k, we obtain $P_{blue} = 6.6 \times 10^3$
\msol \k, $P_{red} = 2.3 \times 10^3$ \msol \k, $E_k^{blue} = 1.4
\times 10^{48}$ erg and $E_k^{red} = 4.2 \times 10^{47}$ erg.
Comparing our results with the work of \citet{wu04} who based on
compiled data concluded that the mass, momentum, and energy of
molecular outflows range from 10$^{-3}$ to 10$^{3}$~\msol, 10$^{-3}$ to
10$^{4}$ \msol \k~and 10$^{38}$ to 10$^{48}$ erg, respectively, we can
infer that in this work we are dealing with massive and energetic
outflows.

Following \citet{curtis10} we define the dynamical time for the blue
and red wings, $t_{dyn}$, as the time for the bow shock travelling at
the maximum velocity in the flow, $V_c$, to travel the projected lobe
length, $R_{lobe}$:

\begin{equation}
t_{dyn}=\frac{R_{lobe}}{V_c}
\end{equation}

\noindent with $R_{lobe}^{blue}$ = 1.8~pc and $R_{lobe}^{red}$ =
1.6~pc. These values were obtained by inspecting the blue and red lobe
sizes in Fig. \ref{blue-red-wings} and considering a distance of
8.3~kpc. We obtain a similar dynamical time of both lobes, $t_{dyn} =
0.8 \times 10^5$~yr. According to \citet{beuther02}, flow ages are
good estimates of protostar lifetimes. We can then assume that the
protostar responsible of the outflow activity has an age of the order
of $10^5$~yr. This is consistent with the detection of ionized gas in
the region, since to be able to ionize its surroundings the age of the
protostar should be at least $10^5$~yr \citep{sri02}.

\subsection{HCO$^{+}$ as tracer of molecular outflows}
\label{hco+}

As an independent test of out findings, we performed a study of the
HCO$^+$ abundance towards G45.47. HCO$^{+}$ is believed to be the
dominant ionized species in dense dark clouds
\citep{rawlings00,dalgarno84}, being very important for the
ion-neutral chemistry. The ionization fraction of molecular clouds is
a relevant parameter to study the cloud chemistry and dynamics, and
hence the star forming processes.  Star formation occurs in the
interior of dense cores, regions of high extinction where
self-shielding prevents the UV photoionization of H$_{2}$. Thus,
cosmic ray ionization is believed to dominate photoionization in dense
cores \citep{mckee89}.  However, when YSOs are formed within the
cores, the shocks and outflows produce deep changes in the chemistry
and photoionization processes are very likely.  Indeed, in star
forming regions occur a molecular enrichment due to the desorption of
molecular-rich ice mantles, followed by photochemical processing by
shock-generated radiation fields.  In particular, towards YSOs which
are driving outflows, it is expectable an enhancement in the HCO$^{+}$
abundance \citep{rawlings00,rawlings04}. In what follows, we estimate,
using a simple chemical network, the HCO$^{+}$ abundance that would be
produced by an standard cosmic ray ionization rate in the G45 clump in
order to compare with the abundance obtained from our observations.

The starting chemical reaction to form HCO$^{+}$ is the production of H$_{3}^{+}$,
mainly formed in the interaction between the cosmic rays (c.r.) and 
the molecular gas (e.g. \citealt{Oka06}):
\begin{eqnarray}
\centering
{\rm H_{2} + c.r.}  &\rightarrow&  {\rm H_{2}^{+} + e^{-} }  \nonumber \\
{\rm H_{2}^{+} + H_{2} }   &\rightarrow&  {\rm H_{3}^{+} + H. }
\label{chem1}
\end{eqnarray}
The rate equation of this reaction is
\begin{equation}
\zeta_{\rm H_{2}}~n({\rm H_{2}}) = \frac{dn({\rm H_{3}^{+}})}{dt},
\label{rate1}
\end{equation}
where $\zeta_{\rm H_{2}}$ is the cosmic ray ionization rate and $n$ the density of
the molecular species.
By considering that the HCO$^{+}$ is mainly formed by the reaction of H$_{3}^{+}$
with CO: 
\begin{equation}
{\rm H_{3}^{+} +  CO \rightarrow  HCO^{+} + H_{2} }
\label{chem2}
\end{equation}
and destroyed through recombination with electrons,
\begin{equation}
{\rm HCO^{+} + e^{-} \rightarrow CO + H,  }
\label{chem3}
\end{equation}
the rate equations can be equated, leading to
\begin{equation}
k_{\rm HCO^{+}}~n({\rm H_{3}^{+}})~n({\rm CO}) = k_{\rm CO}~n({\rm HCO^{+}})~n(e),
\label{rate2}
\end{equation}
where $k_{\rm HCO^{+}}$ and $k_{\rm CO}$ are the coefficient rates and $n(e)$ the
electron density.
Assuming that the main destruction mechanism for H$_{3}^{+}$ is through the
formation of HCO$^{+}$, the rate
of destruction of H$_{3}^{+}$ is therefore equal to the formation rate of HCO$^{+}$;
this implies that the left side of 
equation (\ref{rate1}) is equal to right side of equation (\ref{rate2}). Then, it is
possible to write an expression
for the cosmic ionization rate as a function of the molecular densities:
\begin{equation}
\zeta_{\rm H_{2}} =  k_{\rm CO} \frac{n({\rm HCO^{+}})~n(e)}{n({\rm H_{2}})}.
\label{chem4}
\end{equation}
The rate coefficient $k_{\rm CO}$, extracted from the UMIST database
\citep{woodall07}, is: $k_{\rm CO} = 2.4 \times 10^{-7} (T/300 {\rm K})^{-0.69}$.
Equation (\ref{chem4}) can be approximated using the column densities, leading:
\begin{equation}
\zeta_{\rm H_{2}} \simeq  k_{\rm CO} \frac{{\rm N(HCO^{+})}~X(e)~n({\rm H_{2}})}{\rm
N(H_{2})},
\label{chem5}
\end{equation}
where $X(e)$ is the electron abundance. Using this equation and
assuming typical values for the cosmic ionization rate and electron
abundance of $\zeta_{\rm H_{2}} = (1-5) \times 10^{-17}$ s$^{-1}$
\citep{dalgarno06} and $X(e) \sim 10^{-7}$ \citep{bergin99},
respectively, we derive the HCO$^{+}$ column density. We use
N(H$_{2}$) $= 2 \times 10^{23}$ cm$^{-2}$ and $n({\rm H_{2}}) = 1
\times 10^{5}$ cm$^{-3}$, as derived above, and we assume $T = 20$
K. Finally, we obtain that the N(HCO$^{+}$) should be in the range
$(1.5 - 6.5) \times 10^{14}$ cm$^{-2}$, leading to an abundance
$X({\rm HCO^{+}})$ in between $6.5 \times 10^{-10}$ and $3.2 \times
10^{-9}$.

On the other hand, we analyze the central HCO$^{+}$ J=4--3 spectrum
and use the RADEX code \citep{tak07} to derive its column density.
Assuming background and kinetic temperatures of 2.73 K and 20 K,
respectively, and the same H$_{2}$ density used above, we varied the
HCO$^{+}$ column density until obtaining a good fit for the observed
peak temperature. The best fit was obtained with N(HCO$^{+}$)$ \sim
1.5 \times 10^{15}$ cm$^{-2}$, leading an abundance of $X({\rm
  HCO^{+}}) \sim 7.5 \times 10^{-9}$.  This value, more than twice
greater than the lager value obtained above, might indicate that the
cosmic ray ionization is insufficient to produce the observed
HCO$^{+}$ abundance through the described chemical network.
Therefore, following \citet{rawlings00,rawlings04}, we suggest that
the observed HCO$^{+}$ abundance must be mostly produced by the
outflowing activity in the region. In spite of the uncertainties in
the calculations, this is an independent proof pointing to support an
scenario with outflows in the region.

\subsection{Looking for the driving source of the molecular outflows}

\citet{wil96} reported 5\s~resolution observations of the HCO$^+$
J=1--0 transition towards G45.47 obtained with the OVRO millimeter
array. They identified at least five HCO$^+$ J=1--0 clumps and
suggested that the formation of an OB cluster would be taking place in
the region.  However the authors did not carry out any search of
infrared point sources in the region to confirm that hypothesis.  In
this context, we wonder if the massive molecular outflows observed
towards G45.47 were originated by one or more stars. Based on a
near-infrared analysis, we searched for the driving source candidates
of the massive molecular outflows.

\begin{figure}[h]
\centering \includegraphics[width=7.5cm]{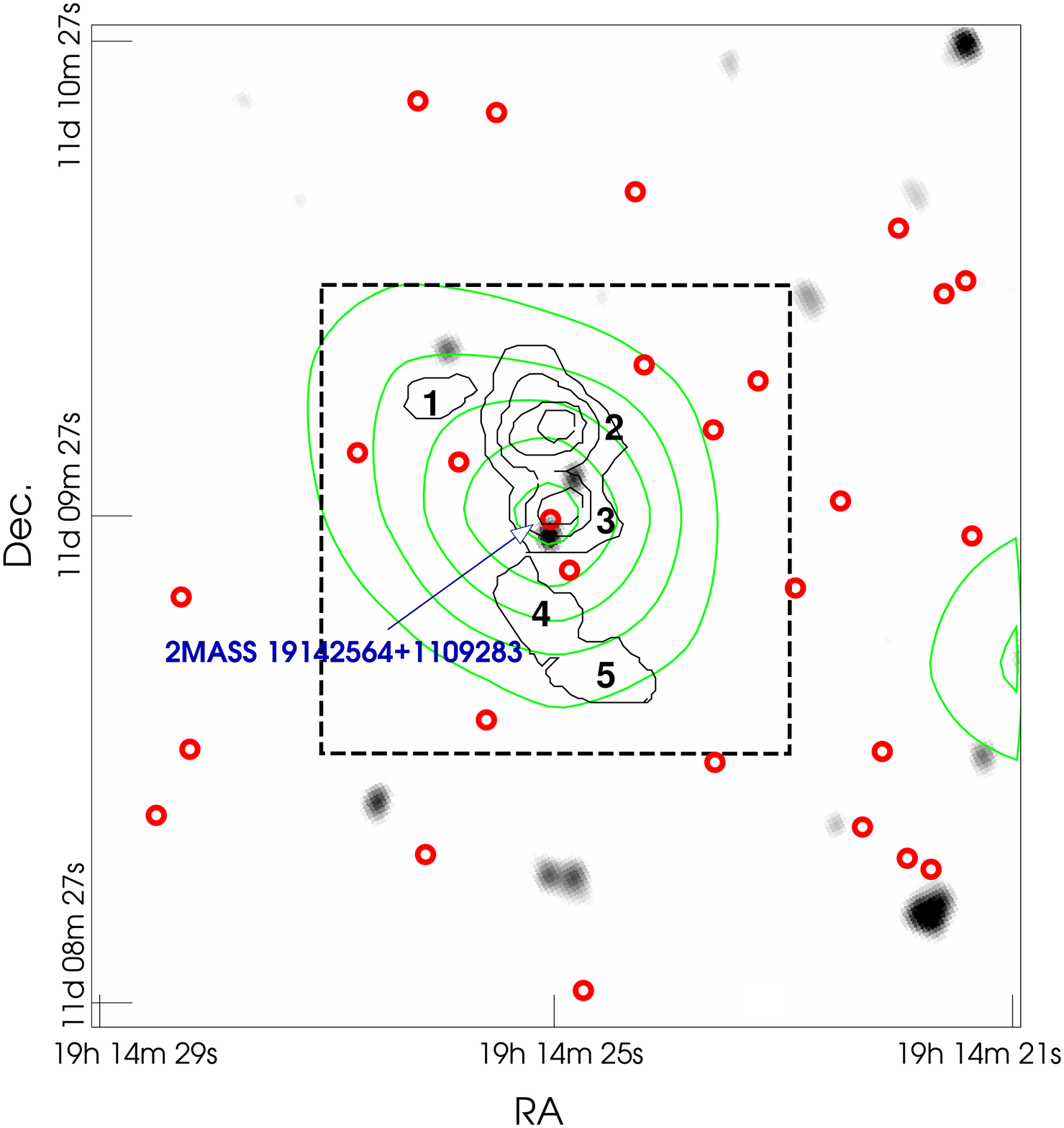}
\caption{Radio continuum emission at 6~cm (MAGPIS; \citealt{white05})
  of the area mapped using ASTE. The green contours represent the
  HCO$^+$ J=4--3 emission distribution integrated between 58 and
  67~\k. The black contours were taken from the paper of \citet{wil96}
  and represent the HCO$^+$ J=1--0 emission distribution integrated
  from 58 and 67~\k. The box indicates the area observed by
  \citet{wil96}. The numbers show the five HCO$^+$ J=1--0 clumps. The
  red circles represent the YSO candidates shown in Fig. \ref{2mass}.}
\label{HCO-clumps}
\end{figure}

\begin{figure}[h]
\centering \includegraphics[width=7.5cm, angle=-90]{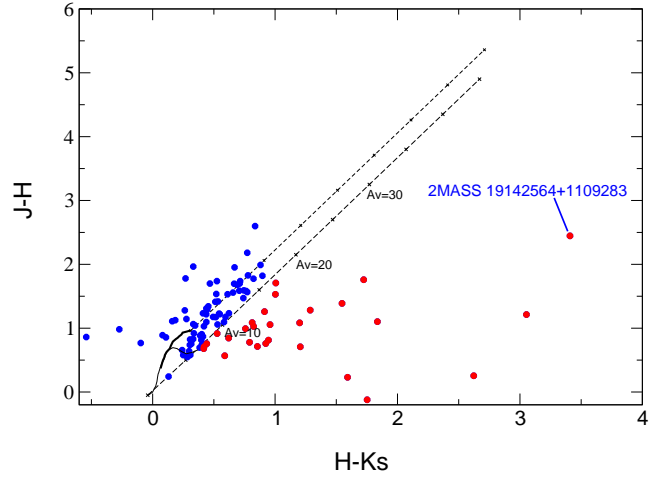}
\caption{Color-color diagram of the 2MASS infrared sources in the
  vicinity of G45.47. The two solid curves represent the location of
  the main sequence (thin line) and the giant stars (thick line)
  derived from \citet{bessell88}. The parallel dashed lines are
  reddening vectors with the crosses placed at intervals corresponding
  to five magnitudes of visual extinction. We have assumed the
  interstellar reddening law of \citet{rieke85} (A$_J$/A$_V$ =0.282;
  A$_H$/A$_V$ =0.175 and A$_K$/A$_V$ =0.112).  The sources reddened by
  circumstellar dust are indicated as red circles.}
\label{2mass}
\end{figure}

Figure \ref{HCO-clumps} shows the radio continuum emission at 6~cm
obtained from The Multi-Array Galactic Plane Imaging Survey (MAGPIS;
\citealt{white05}) of the area mapped using ASTE. The green contours
represent the HCO$^+$ J=4--3 emission distribution integrated between
58 and 67~\k. The black contours were taken from the paper of
\citet{wil96} and represent the HCO$^+$ J=1--0 emission distribution
integrated in the same velocity interval.  The region observed by
\citet{wil96} is indicated by the dashed box.  The five clumps
reported by the authors have been labeled.  The positional coincidence
among the center of the HCO$^+$ J=4--3 clump, the clump 3 of HCO$^+$
J=1--0 and a radio continuum source, which has been identified as the
UCHII region G45.47, is striking.  The red circles indicate the
location of the YSO candidates in the observed region (see red circles
in Fig. \ref{2mass}). Among the five YSO candidates projected onto the
HCO$^+$ J=4--3 emission, 2MASS 19142564+1109283 is the only one having
a positional coincidence with a clump of HCO$^+$ J=1--0 (clump 3). We
do not find any embedded infrared sources in the others four clumps,
suggesting that these clumps could be in a prestellar core stage or
they might be tracing the origin of the molecular outflows. We suggest
that the most likely candidate to be the driving source of the
molecular outflows is 2MASS 19142564+1109283.

 Figure \ref{2mass} shows a color-color diagram (CCD) including all
 the 2MASS point sources located in the observed region. The red
 circles represent the sources with infrared excess (YSO candidates)
 which have been shown in Figure \ref{HCO-clumps}. The reddenest
 object is 2MASS 19142564+1109283. The blue circles represent main
 sequence or giant star candidates.

\begin{figure}[h]
\centering
\includegraphics[width=8cm]{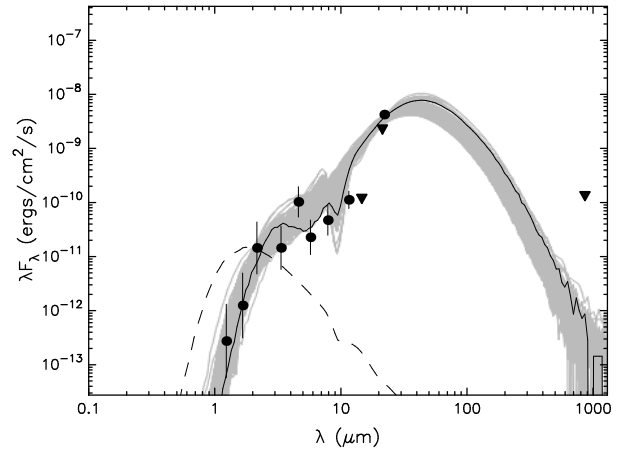}
\caption{Best fitting SED models for G45.47. The fill circles indicate the
  measured fluxes at the {\it JHK} 2MASS bands, {\it Spitzer}-IRAC
  bands at 5.8 and 8.0 $\mu$m, and the WISE bands at 3.4, 4.6, 12 and
  22 $\mu$m. Triangles indicate fluxes considered as upper limits, as
  MSX bands at 14 and 21 $\mu$m and the SCUBA band at 850
  $\mu$m. Black and gray solid curves represent the best-fit model and
  the subsequent good fittings, respectively. The dashed line shows
  the best-fit model for a central source contribution in absence of
  circumstellar dust.}
\label{SED}
\end{figure}

To characterize the infrared source 2MASS 19142564+1109283 we perform
a fitting of its spectral energy distribution (SED) using the tool
developed by
\citet{rob07}\footnote{http://caravan.astro.wisc.edu/protostars/}. We
adopt an interstellar extinction in the line of sight, A$_v$, between 5
and 17 magnitudes. The range of A$_v$ was chosen by inspecting the
location in a [{\it H-K}] vs [{\it J-H}] CCD of the 2MASS sources in
the region (see Fig. \ref{2mass}).  To construct the SED we consider
the fluxes at the {\it JHK} 2MASS bands, {\it Spitzer}-IRAC 5.8 and 8
$\mu$m bands, WISE 3.4, 4.6, 12 and 22 $\mu$m bands, MSX 14 and 21
$\mu$m bands, and SCUBA 850 $\mu$m band.  The fluxes of the datasets
having lower angular resolution (MSX and SCUBA) were considered as
upper limits.  Figure \ref{SED} shows the best fitting SEDs models for
2MASS 19142564+1109283. We select models that satisfies the condition:

\begin{equation}
\chi^2-\chi_{best}^2 < 2N,
\label{selmol}
\end{equation}

\noindent where $\chi_{best}^2$ is the minimum value of the $\chi^2$
among all models, and $N$ is the number of input data fluxes (fluxes
specified as upper limit do not contribute to $N$). Hereafter, we
refer to models satisfying Eq. \ref{selmol} as ``selected models''.

From the SED fitting several model parameters can be obtained and, as
expected, some can be better constrained than others.  Taking into
account the values yielding by the ``selected models'' for several
parameters, we derive the following main results:

\begin{itemize}

\item the total luminosity distribution of the source has an average
  value of $2 \times 10^4$~\lsol~with a spread between $8 \times 10^3$
  and $5 \times 10^4$~\lsol.

\item embedded in the molecular clump there is a massive protostar of
  about 15 \msol~(early B-type star) accreting material from its
  envelope at large rates, $\dot M_{env} \sim
  2\times10^{-4}$~\msol~yr$^{-1}$.

\item despite the large spread in the central source's age
  distribution that goes from $2 \times 10^3$ to $10^6$ yr, it peaks
  at about $3\times 10^5$ yr, as expected for protostars that has
  begun to ionize their surroundings. Besides, this result is in
  agreement with the dynamical time, $t_{dyn} = 1 \times 10^5$~yr,
  derived in Section \ref{outflows}.

\end{itemize}

\section{Summary}

We carried out a study of the UCHII region G045.47+0.05 and its
surroundings based on molecular lines observations and public infrared
data.  We find a molecular gas condensation associated with
G045.47+0.05. The detection of the CS J=7--6 transition reveals that
the UCHII region is still embedded in warm and dense molecular gas.
The CS spectrum obtained towards the position (0, 0) has a pronounced
self-absorption dip at about 59~\k with the red velocity component
more intense than the blue one. This asymmetry in the CS profile
suggests an expansion of the molecular gas. 

All molecular transitions have the same main three velocity components
with a self-absorption dip at about 58-59~\k which correspond to the
central velocity of the molecular cloud GRSMC G045.49+00.04 where
G45.47 is embedded.

Based on the ratio between the $^{12}$CO and $^{13}$CO J=3-2
transitions we estimate the $^{13}$CO opacity, $\tau_{13}$, towards
the center of the clump in about 1.9, revealing that this line is
optically thick.  This result is in agreement with the structure
exhibited by the $^{13}$CO spectrum towards the center of
G045.47+0.05.

Using the RADEX code we derive a $^{13}$CO J=3-2 column density and
H$_2$ volume density of about $2.8 \times 10^{17}$~cm$^{-2}$ and
10$^5$~cm$^{-3}$, respectively. The H$_2$ column density and the total
mass of the clump were estimated in $2.1 \times 10^{23}$~cm$^{-2}$ and
$M \sim 10^4$~\msol, respectively.

From an independent estimate based on the dust continuum emission at
1~mm we derive $N_{gas} \sim 4 \times 10^{23}$~cm$^{-2}$, $M_{gas}
\sim 8520$~\msol, and a volume density, $n$(H$_2) \sim 1.4 \times
10^5$~cm$^{-3}$, in good agreement with the values derived from the
molecular lines.

From the analysis of the $^{12}$CO J=3-2 and HCO$^+$ transitions we
report the presence of molecular outflows related to the UCHII region
G045.47+0.05. We estimate the blue and red lobes masses in about 300
and 120~\msol, respectively. The total mass of the outflows represents
about 4\% of the clump mass. The dynamical time of both lobes, good
indicator of the protostar age, was estimated in about $8 \times
10^4$~yr. This agrees with the presence of ionized gas around the
central object.

Based on infrared data we searched for the protostar(s) that must be
driving the outflows. We find that the source 2MASS 19142564+1109283
is a YSO candidate located onto the center of the clump in coincidence
with the radio continuum emission associated with the UCHII
region. Besides this infrared source is the only YSO candidate related
to one of the five HCO$^+$ clump detected by \citet{wil96}. We did not
detected any YSO candidate associated with the other four HCO$^+$
clumps. This results suggests that the massive molecular outflows are
generated by the protostar 2MASS 19142564+1109283.

Finally, from a spectral energy distribution analysis of the source we
derive a total luminosity, a protostar's mass, and protostar's age of
about $2 \times 10^4$\lsol, 15~\msol (early B-type star), and $3
\times 10^5$~yr, respectively.

Massive molecular outflows give, on large scales, a good piece of
information about the physical processes taking place in the innermost
parts of the star-forming cores. Their detection give indirect
evidence that massive star formation is a scaled-up version of low
mass star formation. The identification of a massive protostar that
would be generating the massive outflows reinforces this scenario.

\section*{Acknowledgments}

We wish to thank the referee, Dr. Herpin, whose constructive criticism
has helped to make this a better paper.  M.O., S.P., S.C., and
G.D. are members of the {\sl Carrera del Investigador Cient\'\i fico}
of CONICET, Argentina. This work was partially supported by Argentina
grants awarded by Universidad de Buenos Aires, CONICET and ANPCYT.
M.R. wishes to acknowledge support from FONDECYT (CHILE) grant
No108033. She is supported by the Chilean Center for Astrophysics
FONDAP No. 15010003. The ASTE project is driven by Nobeyama Radio
Observatory (NRO), a branch of National Astronomical Observatory of
Japan (NAOJ), in collaboration with University of Chile, and Japanese
institutes including University of Tokyo, Nagoya University, Osaka
Prefecture University, Ibaraki University, Hokkaido University and
Joetsu University of Education.

\bibliographystyle{aa}  
\bibliography{biblio}
\IfFileExists{\jobname.bbl}{}
{\typeout{}
\typeout{****************************************************}
\typeout{****************************************************}
\typeout{** Please run "bibtex \jobname" to optain}
\typeout{** the bibliography and then re-run LaTeX}
\typeout{** twice to fix the references!}
\typeout{****************************************************}
\typeout{****************************************************}
\typeout{}
}

\label{lastpage}
\end{document}